# Understanding of the role of magnetic fields: Galactic perspective

A. Lazarian, S. Boldyrev, C. Forest, J. Sarff, P. Terry
Universitty of Wisconsin-Madison

White Paper submitted to the Galactic Neighborhood (GAN) Panel.

## I. Magnetic Fields in the Galactic Neighborhood

**Goal of White Paper**: A combination of observation, theory, modeling, and laboratory plasma experiments provides a multifaceted approach to develop a much greater understanding of how magnetic fields arise in galactic settings and how these magnetic fields mediate important processes that affect the dynamics, distribution, and composition of galactic plasmas. An important emphasis below is the opportunity to connect laboratory experiments to astrophysics. This approach is especially compelling for the galactic neighborhood, where the distribution and character of magnetic fields can be observed with greater detail than what is possible elsewhere in the universe. The ability to produce laboratory plasmas with unparalleled accessibility permits an even greater level of detail to be assessed and exposed. Theory and modeling provide fundamental ways to understand important processes, and they act as the bridge to connect experimental validation to astronomical observations. In many cases the studies that utilize this approach can make use of existing laboratory facilities, resulting in a cost that is quite small compared to the cost of measurements in dedicated space missions.

**Importance of magnetic fields:** Magnetic fields are ubiquitous in space, and their role in affecting astrophysical processes is difficult to overestimate. For instance, they are known to be important in the acceleration of cosmic rays, star formation (because their pressure often prevents cloud collapse), and MHD heating of the interstellar medium. To understand these cosmic fields, we must, first of all, understand their dynamics in turbulent plasmas ("magnetic turbulence"), the dynamics of magnetic field amplification (usually called "dynamo"), and topology changes ("reconnection"). Complex interaction of magnetic fields with plasmas and cosmic rays, and the back-reaction of these processes on the magnetic field dynamics and structure, makes these processes difficult to describe quantitatively. Therefore, the Galactic Neighborhood provides an ideal testing ground where high resolution observations can be compared with theoretical modeling, numerical simulations and laboratory experiment. In the text below we discuss magnetic processes that span 12 orders of magnitude of scales.

The major questions relevant to this topic are: "*How are magnetic fields created at the Galactic scale? How do they mediate the interstellar dynamics over a tremendous range of scales, from 1000pc to 100km? How is magnetic energy irreversibly converted into heat and fast particles?*"

**Observational progress**: New instruments that are going to be available in the next decade, e.g. SKA, LOFAR, ALMA can provide much better quantitative insight into magnetic field properties and structure. Detailed polarimetry surveys as well as new techniques of studying magnetic field, e.g. through Hanle effect and atomic alignment, expand further the prospects of the observational studies. Thus the time will be ripe for detailed comparisons of observations with theoretical expectations.

**Organization of the paper:** For clarity in discussion, we separate the discussion of magnetic fields according to their scale: (1) largest scale (section II), (2) meso or intermediate (section III), and (3) small scale (section IV) usually associated with dissipation mechanisms. For each of these there are important questions and identifiable processes. Though the scale separation is convenient for discussion, it is important to recognize that all scales are connected. Indeed, the role of magnetic fields in mediating matter and energy at different scales is one of the major points we highlight. We also stress the importance of understanding properties of magnetic fields in the Galactic Neighborhood for quantitative modeling of extragalactic processes and reliable separation of CMB and foreground polarization (section V). A short summary is provided in section VI. Sections II, III, and IV contains sub-questions which are necessary to address in answering the major questions above.

**II. Large scale Field**
A magnetic field is present on the scale of the Galaxy[1] ($10^3$ parsec scale). Its origin is usually attributed to the galactic dynamo, the operation of which is poorly understood. Unfortunately, the traditional textbook mean field dynamo theory does not conserve magnetic helicity, contrary to observations in laboratory plasma experiments and theoretical expectations[2]. In addition, the inverse cascade of magnetic helicity in a turbulent plasma is very much influenced by the correlations between the velocity and magnetic field fluctuations, which are inherently weak within the accepted treatments of dynamo.

The key questions here are: "*What is the correct formulation of dynamo theory? What is the role of velocity and magnetic field correlations for magnetic field generation?*"

Laboratory studies provide a new outlook on the problem. For instance, spontaneous yet robust magnetic flux conversion (a central component of the dynamo process) observed in laboratory fusion research plasmas like the reversed field pinch suggest that large-scale dynamo can be generated by instabilities in which the magnetic and velocity fluctuations are dynamically linked[3]. Using an array of sophisticated measurements, laboratory fusion experiments offer great opportunity to test new ways of approaching the dynamo. Advanced computations offer the opportunity to explore further the inherently nonlinear mechanisms.

Unfortunately, there has been almost no investigation of laboratory plasmas with relatively high beta and large plasma flows so that the system is kinetic-energy dominated. For instance, the strongly magnetized laboratory fusion plasmas like the reversed field pinch are magnetic-energy dominated. However, high beta plasmas with large plasma flows are experimentally accessible, and would provide a first-time opportunity to investigate astrophysically relevant plasma conditions in which the kinetic energy dominates. In fact, one possible approach uses a two-vortex flow generated by oppositely directed rotation in the upper and lower hemispheres of a spherical confinement chamber. This flow is similar to the two vortex flow first described in Dudley and James[4], which has been shown to self-excite above a critical magnetic Reynolds number > 350. In addition, a convection driven alpha-omega dynamo experiment is feasible. Differential rotation can efficiently generate a strong toroidal field from a relatively weak poloidal field through the omega effect. Helical turbulence can be generated by injecting light ions into a strongly rotating, heavy ion flow. The turbulence is expected to regenerate the poloidal field from the toroidal field and thereby close the feedback loop required for self-excitation. This suggests, that if the laboratory studies above are combined with numerical and analytical efforts, a new approach to the magnetic field generation can be developed.

Apart from conserving helicity, the galactic dynamo requires changes of the magnetic field topology that are enabled by reconnection. Recent years have been marked by remarkable progress in simulating fast collisionless reconnection[5]. This type of reconnection is restricted, however, to situations where the length of the current sheet is no more than several times larger than the electron mean free path . This regime is not applicable to interstellar gas, for which the reconnection is collisional[6]. If collisionless reconnection is the only type of fast reconnection, this means that most of the interstellar MHD codes do not correctly represent astrophysical reality! Indeed, the inevitable numerical diffusivity in the codes makes reconnection generically fast. This potential difference not only raises serious questions about simulations of the galactic dynamo, but also the entire crop of simulations of interstellar medium dynamics, star formation, etc. In fact, the absence of codes to properly simulate interstellar magnetic fields, might urgently require starting numerical interstellar research from scratch, e.g., first via creation of a new generation of codes that explicitly forbid reconnection. Whether most of the current interstellar simulations are astrophysically irrelevant depends on whether a fast reconnection process in collisional environments will be proven to operate.

The outstanding questions here are: "*How can reconnection proceed in collisional environments? Can turbulence accelerate reconnection? How does partial ionization of interstellar plasma affect reconnection speeds? What are the consequences of magnetic reconnection for key interstellar processes?*" These burning questions can and must be addressed through the synergy of the theory with laboratory and numerical experiments. Incidentally, this does not mean attempts to exactly reproduce astrophysical conditions in the laboratory, but rescaling plasma characteristics in the case of laboratory research and testing theoretical scalings for the regimes accessible numerically. The consequences of such studies, e.g., predictions of cosmic ray acceleration if magnetic reconnection is fast, should be tested observationally.

### III. Meso and Intermediate Scales
The interstellar medium (ISM) supports a remarkable self-organizing process in the form of a turbulent energy cascade observed through the fluctuation spectrum of electron density, fluctuating component of the magnetic field, and fluctuations of Doppler shifted emission/absorption lines. The extraordinarily broad range of scales encompassed by the cascade, from dozens of parsecs to hundreds of kilometers[7], makes it a potent participant and regulator in a variety of processes. It is a conduit of magnetic energy across these scales, linking large scale stirring from super novae, stellar winds and large-scale instabilities. At the smallest scales it is thermalized by dissipative processes that are likely to be common to ion heating in the solar corona, to anisotropic pressures in the solar wind, reconnection, etc.

Magnetic turbulence is important at all scales and in all phases of the interstellar medium. It plays a key role in generating and structuring large-scale magnetic fields that help to trap, scatter and accelerate cosmic rays. It modifies large-scale instabilities, thus affecting dynamics of galactic matter. It mediates energy transfer from large Galactic scales to small dissipation scales via a turbulent cascade, thus affecting interstellar plasma heating. It mixes and homogenizes the interstellar gas, which is intermittently "polluted" by supernovae, thus affecting the interstellar chemistry. It plays a crucial role in density fragmentation in cold molecular clouds, thus mediating the process of star formation[8].

Crucial aspects of the physics of the cascade above are not well understood and hotly debated. The problem here is that the commonly used brute force numerical approach has serious limitations. First of all, if reconnection in the interstellar medium happens to be slow, there is a possibility (which has not yet been convincingly refuted) that all simulation related to the interstellar medium, including those of turbulence have significnant errors due to a dramatic disparity between the rates of astrophysical and numerical reconnection. If the reconnection rates in collisional environments are proven to be fast, there still exists a problem of the enormous disparity of the magnetic Reynolds numbers (Rm) in simulations and interstellar gas. The aforementioned Rm, which characterize the degree of frozenness of magnetic field within eddies, may differ by a factor larger than $10^{10}$, which, naturally, calls for caution while interpreting numerical results. In addition, the turbulence takes place in a multiphase medium, where observed interstellar structures cover not only a large range of scales but also a large range of densities and temperatures. On top of all this, interstellar turbulence has spatially localized sources and sinks, which makes magnetic wave disturbances imbalanced in terms of the magnitudes of oppositely propagating wave components.

Traditionally, analytic and numerical studies of magnetic turbulence have concentrated on the homogeneous and incompressible cases. However, astrophysical plasmas indicate an urgent necessity for addressing much broader regimes. Among the most relevant problems are the turbulent generation of magnetic fields (as observations indicate that magnetic energy can sometimes be below the equipartition energy), the description of compressible turbulence (as compressibility is very important for the interstellar medium), the study of turbulence with a different ratio of magnetic to thermal energy (as this ratio varies for different interstellar phases), the study of turbulence with a strong and weak guide field (as the ratio of random to regular magnetic field vary in the ISM), the study of imbalanced turbulence where energy fluxes in opposite directions along the guide field are non-equal (as expected in the presence of sources of turbulence), the study of turbulence in partially ionized gas (as ISM hosts a lot of neutral atoms), etc.

The basic questions of astrophysical interest include: *What is the turbulence spectrum and its anisotropy? How strong is the coupling of the fundamental MHD modes, i.e. slow, fast and Alfven modes in the cascade? How fast does the turbulence decay?* These questions should be addressed for the different regimes above.

The complexity of the problem calls for a synergetic approach combining concerted and coordinated effort involving observations of the local ISM, measurements in related laboratory plasmas, development of theory and supporting numerical computation[α]. The importance of laboratory studies for understanding astrophysical turbulence is frequently underestimated. Consider Fig. 1, which reveals similarities between the magnetic fluctuation spectra from solar wind observations and from the Madison Symmetric Torus (MST)[3]. Both spectra show magnetic fluctuation spectra with apparent power-law-like falloff, with putative knees in the rough vicinity of gyroradius scales, and both have been argued to be subject to cyclotron resonance dissipation. Laboratory plasmas like MST offer unexcelled diagnostic access relative to astrophysical plasmas, incorporating techniques based on laser scattering, beam emission, Thompson scattering, and spectroscopy. Some methodologies and technologies, such as Faraday rotation based polarimetry, are shared with astrophysical observation, allowing direct interchange of advances. MST is just one example of laboratory opportunities.

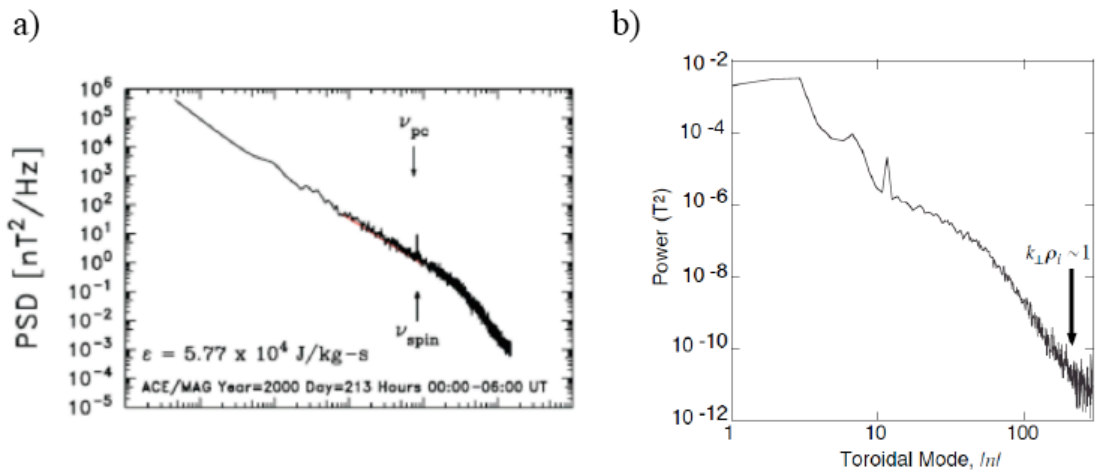

Fig. 1 a) Spectrum of solar wind turbulence from Smith et al. b) Spectrum of magnetic turbulence in the Madison Symmetric Torus.

**IV. Dissipation Scales and Microscales**

Observations indicate that small-scale density structures (from hundred to several AU) are present in interstellar gas[9]. This is indicative of magnetic self-organization taking place at the scales approaching the turbulence dissipation scale. Solar flares accompanied by acceleration of energetic particles are examples of magnetic reconnection and transfer of magnetic energy to particle acceleration. At the same time, solar corona observations indicate heating of the heavy element ions ("minor ions"), which are believed to be a consequence of the magnetic turbulence dissipation[10].

The self-similar inertial range of the turbulence that originates at the injection turbulence scale is truncated at the dissipation scale[β]. The latter scale depends on the nature of the

---

[α] While we stressed the problems of the brute force approach, numerics is indispensable for testing theoretical predictions/scalings for the range of Reynolds and magnetic Reynolds numbers accessible by modern computers.

cascade and the properties of the particular astrophysical environment. For instance, in highly ionized interstellar medium magnetosonic fast modes can be damped through collisionless damping, while Alfven modes can propagate to smaller scales. Similarly, the dissipation of magnetic turbulence may vary in the presence cosmic rays, neutral atoms, minority ions, charged dust etc.  As turbulence dissipates it heats the medium, transfers its energy to energetic particles, induces various plasma effects. Understanding of the key processes in most cases requires going beyond the MHD treatment of the fluid. For instance, the redistribution of heating between protons and electrons may depend on the anisotropy of whistler turbulence [11].

Moreover, the dissipation of turbulence can happen intermittently through the volume. Depending on the degree of intermittency the consequences of turbulence may vary dramatically. For instance, Falgarone et al.[12] showed that intermittent turbulence can dramatically alter interstellar chemistry, via exothermal reactions induced by very concentrated vortices.  Whether or not such a radical revision of interstellar chemistry paradigm is necessary depends on the degree of the actual turbulence intermittency.

The relevant questions are: *How does MHD turbulence dissipate? How does the energy get distributed between charge carriers of different mass? How high is the intermittency of energy dissipation?*

Also, gyro-kinetic simulations of the turbulent dissipation in plasma, two fluid simulations of turbulent dissipation in partially ionized gas open wide avenues for testing properties of magnetic fields on small scales numerically.  In addition, both in situ magnetic field measurements and observations of the solar corona can be another vital source of information. However, the role of laboratory experiments should not be underestimated.  For instance, MST turbulence (see Figure 1) shows evidence for the type of magnetic wave activity that govern magnetic turbulence in the interstellar medium as turbulence enters scales where electron compressibility effects arise, and is associated with ion heating.

**V. Broad Impact of Better Understanding of Magnetic Fields**
The synergetic approach described above for studies of magnetic fields in the Galactic Neighborhood can clarify the role that magnetic fields play in the dynamics and physics of fundamental astrophysical processes, e.g., transport of heat and cosmic rays, star formation, accretion. This should help quantify the effect of astrophysical magnetic field in extragalactic systems for which high-resolution observations are not readily available.

The studies of the properties of magnetic fields in the Galactic neighborhood should provide important input for realistic parametrization of key extragalactic astrophysical processes. For instance, if the role of magnetic field and magnetic turbulence in star formation is clarified, this will boost the reliability of extragalactic simulations that require the star formation rates as their input. Similarly, understanding of magnetic

---

β The degree of self-similarity in the inertial range is debated issue that requires further investigation, however.

reconnection should shed light on many fundamental astrophysical processes, including accretion of matter. Last, but not the least, understanding of magnetic field structure in the Galactic neighborhood is important to understand the properties of polarized foregrounds that interfere with the CMB polarization.

**VI. Summary: Towards Quantitative Understanding of Magnetic Fields**
Magnetic fields on different scales are interrelated and interdependent. For instance, the structure of magnetic field which is governed by the galactic large scale dynamo depends on reconnection, which may depend either on the properties of magnetic turbulence at the mesoscales or even on properties of plasma at microscales comparable with the proton gyroradius. The advocated synergetic approach includes the traditional interplay between experiment, observations, and theory in developing a conceptual picture and understanding; the validation of numerical algorithms by appropriate comparison with experiment; and the application of validated codes to astrophysical phenomena. The later allows gaps in parameter regimes between laboratory and astrophysical plasmas to be bridged, through analysis and synthesis of common physics.

Several recent developments make this problem ripe as a high priority effort. Magnetic fields in near earth galactic environment are readily accessible and observable via dispersion techniques, rotation measure, polarimetry, spectroscopy, etc. In addition, the wealth of spectroscopic surveys provides complementary input on the velocity fluctuations, which carry essential signatures of interstellar turbulence. Moreover, there is the opportunity of synthesis from observations of the solar wind, solar corona, and magnetospheric and auroral plasmas.

As a result of implementing of this approach the fundamental properties of magnetic fields, including their generation by dynamo and turbulence, their dynamical impact on galactic processes, which heavily depends on the properties of turbulence and the reconnection speed, their interaction with cosmic rays, and, finally, production of energetic particles and heating, can be substantially clarified. This progress will have broad impact, as magnetic fields are not confined to the Galactic Neighborhood but are really ubiquitous in Astrophysics.